\documentclass[prl,twoside,preprintnumbers,superscriptaddress,twocolumn,nofootinbib]{revtex4}

\usepackage{amsmath,graphicx,slashed,multirow,color,ulem}
\usepackage{graphicx,epsfig,amssymb,bm,slashed,wasysym,multirow}
\def \beq{\begin{equation}}
\def \eeq{\end{equation}}
\def \beqa{\begin{eqnarray}}
\def \eeqa{\end{eqnarray}}

\begin{document}

\preprint{CERN-PH-TH/2015-108}
\preprint{DCPT/15/28}
\preprint{IPPP/15/14}
\preprint{OUTP-15-10P}

\title{Beauty-quark and charm-quark pair production asymmetries at LHCb}

\author{Rhorry Gauld}
\email{rhorry.gauld@durham.ac.uk}
\affiliation{Institute for Particle Physics Phenomenology, University of Durham, DH1 3LE Durham, United Kingdom}
\author{Ulrich Haisch}
\email{ulrich.haisch@physics.ox.ac.uk}
\affiliation{Rudolf Peierls Centre for Theoretical Physics, University of Oxford, OX1 3PN Oxford, United Kingdom}
\affiliation{CERN Theory Division, CH-1211 Geneva 23, Switzerland}
\author{Ben D. Pecjak}
\email{ben.pecjak@durham.ac.uk}
\affiliation{Institute for Particle Physics Phenomenology, University of Durham, DH1 3LE Durham, United Kingdom}
\author{Emanuele Re}
\email{Emanuele.Re@physics.ox.ac.uk}
\affiliation{Rudolf Peierls Centre for Theoretical Physics, University of Oxford, OX1 3PN Oxford, United Kingdom}

\date{\today}


\begin{abstract} 
The LHCb collaboration has recently performed a first measurement of the angular production asymmetry in the distribution of beauty quarks and anti-quarks at a hadron collider. We calculate the corresponding standard model prediction for this asymmetry at fixed-order in perturbation theory. Our results show good agreement with the data, which is provided differentially for three bins in the invariant mass of the $b \bar b$ system. We also present similar predictions for both beauty-quark and charm-quark final states within the LHCb acceptance for a collision energy of $\sqrt{s} = 13 \, {\rm TeV}$. We finally point out that a measurement of the ratio of the $b \bar b$  and $c \bar c$ cross sections may be useful for experimentally validating charm-tagging efficiencies.
\end{abstract}

\maketitle

\section{Introduction}

The LHCb collaboration has recently performed a first measurement of the angular asymmetry in $b \bar{b}$ production at a hadron collider~\cite{Aaij:2014ywa}. More specifically, LHCb has measured the forward-central asymmetry of $b$-quark pairs~($A_{\rm FC}^{b \bar b}$) with $1 \, {\rm fb}^{-1}$ of Run-I data, collected at a centre-of-mass energy ($\sqrt{s}$) of $7 \, {\rm TeV}$ in $pp$ collisions.\footnote{Subsequently, the forward-backward asymmetry of $b$-quark pairs has also been measured in $p\bar p$ collisions at the Tevatron~\cite{Abazov:2014ysa,Aaltonen:2015mba}.}

Instrumented in the forward region, the LHCb detector operates in a kinematic regime which is well suited to measure heavy-quark production asymmetries. Forwardly produced heavy-quark pairs, particularly at high invariant mass, are sensitive to colliding partons at both moderate and large momentum fractions within the proton. This kinematic sensitivity provides a unique opportunity to perform asymmetry measurements as the dilution from the otherwise overwhelming symmetric gluon-gluon-fusion ($gg$) production mechanism is reduced to a manageable level. Moreover, the ability of the LHCb detector to efficiently tag semi-leptonic $B$ decays has made this measurement possible with the available data set.

A notable feature of $A_{\rm FC}^{b \bar b}$ is that it receives a large correction from purely electroweak (EW) effects when the invariant mass of the $b$-quark pairs is close to the $Z$-boson resonance, as pointed out in~\cite{Grinstein:2013mia}. Measurements of~$A_{\rm FC}^{b \bar b}$ are therefore not only of general importance as a test of the standard model (SM), but also provide valuable model-building input \cite{Grinstein:2013mia,Murphy:2015cha,Bai:2011ed,Saha:2011wr,Strassler:2011vr,Kahawala:2011sm,Drobnak:2012cz,Delaunay:2012kf,Ipek:2013zi}, serving to restrict the set of new-physics scenarios which were suggested as an explanation of the anomalously large forward-backward asymmetry in top-quark pair production as observed at the Tevatron~\cite{Aaltonen:2012it,Abazov:2013wxa,Aaltonen:2013vaf,Abazov:2014oea}. Although the tensions between the experimentally observed top-quark asymmetries and the corresponding SM predictions have been to a large extent resolved, owing to experimental improvements \cite{Abazov:2014cca} and the inclusion of next-to-next-to-leading order~(NNLO) QCD corrections~\cite{Czakon:2014xsa}, measurements of~$A_{\rm FC}^{b \bar b}$ remain interesting in view of the persistent discrepancy between the experimental data and the SM predictions for the $Z \to b \bar b$  pseudo-observables \cite{ALEPH:2005ab}.

In this letter we report on the SM calculation of $A_{\rm FC}^{b \bar b}$, comparing our results to the LHCb data. We also provide predictions for $A_{\rm FC}^{b \bar b}$ and the corresponding charm-quark pair production asymmetry $A_{\rm FC}^{c \bar c}$ at $\sqrt{s} = 13 \, {\rm TeV}$, relevant for data taking commencing in the next two years. Our SM calculations improve on work very recently presented in~\cite{Murphy:2015cha}.

\section{Anatomy of asymmetry}

The forward-central asymmetry in heavy-quark production is defined in terms of the $p p \to Q \bar Q$ cross section~$\sigma$ in the following way 
\beq \label{eq:1}
A_{\rm FC}^{Q \bar Q} = \frac{\sigma \left (\Delta y > 0 \right ) - \sigma \left (\Delta y< 0 \right )}{\sigma \left (\Delta y > 0 \right ) + \sigma\left (\Delta y< 0 \right )} \,,
\eeq
where $\Delta y = |y_Q| - |y_{\bar{Q}}|$ is the difference of the absolute rapidities of the heavy quark $Q$ and anti-quark~$\bar Q$, evaluated on an event-by-event basis. At the LHC, the~$pp$ initial state is symmetric with respect to the direction of the incoming proton. However, an asymmetry is present in the momentum fraction distributions of quark and anti-quarks within the proton as a consequence of the valence content, and the definition of $A_{\rm FC}^{Q \bar Q}$ is chosen to reflect this  ---~i.e.~to gain sensitivity to the underlying parton distribution function (PDF) asymmetry  in the incoming quark momenta $|p_z(q)|-|p_z(\bar{q})|$. Consequently, the asymmetric contribution to the numerator of (\ref{eq:1}) arises from subprocesses of the form $q\bar q \rightarrow Q \bar Q X$, where $X$ denotes a particular final state such as $g$ or~$\gamma$. Note that by crossing symmetry, the subprocesses $q(\bar q)X \rightarrow Q \bar Q q(\bar q)$ also contribute to the numerator. As demonstrated first in~\cite{Kuhn:1998jr,Kuhn:1998kw}, the dominant contribution to the top-quark asymmetry arises at next-to-leading order (NLO) in QCD. The contributions from mixed QCD-electroweak corrections, which are structurally similar to the NLO QCD effects, and pure EW contributions,  which arise at leading order~(LO) due to the presence of both vector and axial-vector couplings in  the subprocess $q\bar{q} \to \gamma/Z \to Q\bar{Q}$, have also been computed~\cite{Kuhn:1998jr,Kuhn:1998kw,Kuhn:2011ri,Hollik:2011ps,Bernreuther:2012sx,Aguilar-Saavedra:2014vta,Gauld:2014pxa}. In the case of $t \bar t$ production at LHCb, the mixed QCD-EW contributions amount to ${\cal O} (10\%)$ of the total QCD corrections, while pure LO EW effects are negligibly small. In contrast, both the $b \bar b$   and $c \bar c$ asymmetries can receive large contributions from pure EW effects when the invariant mass $m_{Q\bar Q}$ of the heavy-quark pair is in the vicinity of the $Z$ pole. In this kinematic regime, higher-order corrections to the LO EW contribution can in principle also become important, and should be included if possible.

The prediction for $A_{\rm FC}^{Q \bar Q}$ is cast in terms of a Taylor series expansion in powers of the strong ($\alpha_s$) and the electromagnetic ($\alpha$) couplings in the following way
\begin{eqnarray} \label{eq:2}
A_{\rm FC}^{Q \bar Q} = \frac{\alpha_s^3 \hspace{0.25mm} \sigma_a^{s (0)}  + \alpha_s^2 \alpha \hspace{0.25mm} \sigma_a^{se (0)} + \alpha^2 \left( {\sigma}_a^{e(0)} +  \alpha_s {\sigma}_a^{e(1)}\right)}{\alpha_s^2 \left(\sigma_s^{s(0)} + \alpha_s \sigma_s^{s(1)}\right) +  \alpha^2 \left(\sigma_s^{e(0)} + \alpha_s \sigma_s^{e(1)}\right)} \, . \hspace{1.5mm}
\end{eqnarray}
Here the terms $\sigma_a^{s (0)}$ and $\sigma_a^{se (0)}$ correspond to the asymmetric NLO QCD and mixed NLO QCD-EW contributions, respectively, while ${\sigma}_a^{e(0)}$ and ${\sigma}_a^{e(1)}$ represent the  pure EW asymmetric contributions and the corresponding leading QCD correction. In the denominator, our calculations include the LO symmetric QCD and pure EW  contributions ${\sigma}_s^{s(0)}$ and~${\sigma}_s^{e(0)}$, as well as the associated QCD corrections ${\sigma}_s^{s(1)}$ and~${\sigma}_s^{e(1)}$. 

Analytic formulas for the term~$\sigma_a^{s (0)}$ can be found in~\cite{Kuhn:1998kw}. Approximate results for the contribution $\sigma_a^{se (0)}$ are also provided in that article, but these results are not applicable in the resonant region~$m_{Q \bar Q }\simeq m_Z$, which is relevant for the beauty-quark and the charm-quark asymmetries. We have therefore computed the relevant formula for $\sigma_a^{se (0)}$ with the help of \texttt{FeynArts}~\cite{Hahn:2000kx} and \texttt{FormCalc}~\cite{Hahn:1998yk}. The numerical integration of these formulas is performed using the \texttt{Vegas} algorithm as implemented in the \texttt{Cuba} library~\cite{Hahn:2004fe}, and the complex scalar one-loop integrals are evaluated with the \texttt{OneLOop} package~\cite{vanHameren:2010cp}. The contributions $\sigma_{a,s}^{e (0)}$ and  $\sigma_{a,s}^{e (1)}$ have been calculated utilising the helicity amplitudes of~\cite{Alioli:2008gx} which have been extended to include the ${\cal O} (\alpha_s)$ corrections associated to the final-state heavy-quark lines.  All of the aforementioned terms have been calculated with physical heavy-quark masses, with the exception of the term $\sigma_{a,s}^{e (1)}$ which is computed in the massless limit $m_Q = 0$.
The final contribution $\sigma_s^{s (1)}$ is obtained with the matrix elements of~\cite{Nason:1987xz}, that are incorporated in \texttt{POWHEG}~\cite{Frixione:2007nw}. To combine all of our predictions, we perform a change of renormalisation scheme for the symmetric QCD computation which is performed in a fixed-flavour-number scheme (the four-flavour scheme for $b$-quark pair production). This is done following the procedure outlined in~\cite{Cacciari:1998it}, allowing the contributions from all subprocesses for both beauty and charm predictions to be consistently convoluted with five-flavour PDFs. We note that several cross checks of our predictions were performed with \texttt{MCFM}~\cite{MCFM}. Further details on our computations, including analytic formulas for all asymmetric terms will be presented elsewhere~\cite{inpreparation}.

Our numerical results are obtained for the following choice of input
parameters~\cite{Agashe:2014kda}: $m_t = 173.25 \, {\rm GeV}$, 
$m_b =
4.75 \, {\rm GeV}$, $m_c = 1.5 \, {\rm GeV}$, $m_W = 80.385 \, {\rm
  GeV}$, $\Gamma_W = 2.085 \, {\rm GeV}$, $m_Z = 91.1876 \, {\rm
  ~GeV}$, $\Gamma_Z = 2.4952 \rm {~GeV}$ and $G_F = 1.16638 \cdot
10^{-5} \, {\rm GeV}^{-2}$. To describe the $Z$ resonance, we adopt
the complex-mass scheme~(see~e.g.~\cite{Denner:1999gp}) and determine
the sine of the weak mixing angle and the electromagnetic coupling
from $s^2_w = 1 - m_W^2/m_Z^2$ and $\alpha = \sqrt{2}/\pi
\hspace{0.25mm} G_F \hspace{0.25mm} m_W^2 \hspace{0.25mm} s^2_w$,
respectively.  All contributions to~(\ref{eq:2}) are computed with
\texttt{NNPDF2.3} NLO PDFs~\cite{Ball:2012cx} using $\alpha_s(m_Z) =
0.119$. Finally, we evaluate the ratio in~(\ref{eq:2}) exactly as written,
rather than performing a further expansion of higher-order terms in 
the denominator.  These are our most advanced theoretical predictions, and 
we will refer to the results obtained in this way as  ``NLO'' asymmetry predictions.
In addition, to explore the convergence of the perturbative series, we also compute 
the leading contribution to the asymmetry with \texttt{NNPDF2.3} LO PDFs and 
$\alpha_s(m_Z) = 0.119$. This corresponds to dropping the terms $\sigma_s^{s(1)}$ 
and $\sigma_{a,s}^{e(1)}$ in the expression~(\ref{eq:2}), and results obtained in this way
  are therefore referred to as ``LO'' asymmetry predictions. 
For both LO and NLO predictions, a scale uncertainty is evaluated by
independently varying the factorisation $\mu_F$ and renormalisation
$\mu_R$ scales by a factor of two around $m_Z$, imposing the
constraint $1/2 < \mu_F/\mu_R < 2$. The total uncertainty of the
predictions for the asymmetry is then found from the envelope of the
different results.

\section{\boldmath Comparison with $\sqrt{s} = 7 \, {\rm TeV}$ data} \label{7TeV}

To compare our predictions to the available data, a fixed-order analysis\footnote{We have investigated resumming potentially large corrections arising from the soft and small-mass logarithms associated with energetic heavy-quark production as calculated in~\cite{Ferroglia:2012ku}, but it is not clear that results obtained in the soft-gluon emission limit are applicable in the case considered here. In consequence, we do not include such effects in our analysis.} is performed which includes the appropriate experimental cuts to mimic the LHCb selection requirements. Jets are clustered using the anti-$k_t$ algorithm \cite{Cacciari:2008gp} with a distance parameter $R = 0.7$. Since in the data the reconstructed jets are corrected to the parton level, this procedure should allow for a fairly good comparison. The reconstructed jets are required to be within the pseudo-rapidity range $2 < \eta < 4$, to have a minimum transverse energy of $E_T > 20 \, {\rm GeV}$ and are also constrained to have an opening angle of $\Delta \phi > 2.6$ in the transverse plane. The calculation of $A_{\rm FC}^{b \bar b}$ is then performed differentially in the invariant mass bins $m_{b\bar b} \in [40, 75] \, {\rm GeV}$, $[75, 105] \, {\rm GeV}$ and $m_{b\bar b} > 105 \, {\rm GeV}$ to match the LHCb analysis. 

\begin{figure}[t!]
\begin{center}
\makebox{\includegraphics[width=0.95\columnwidth,angle=0]{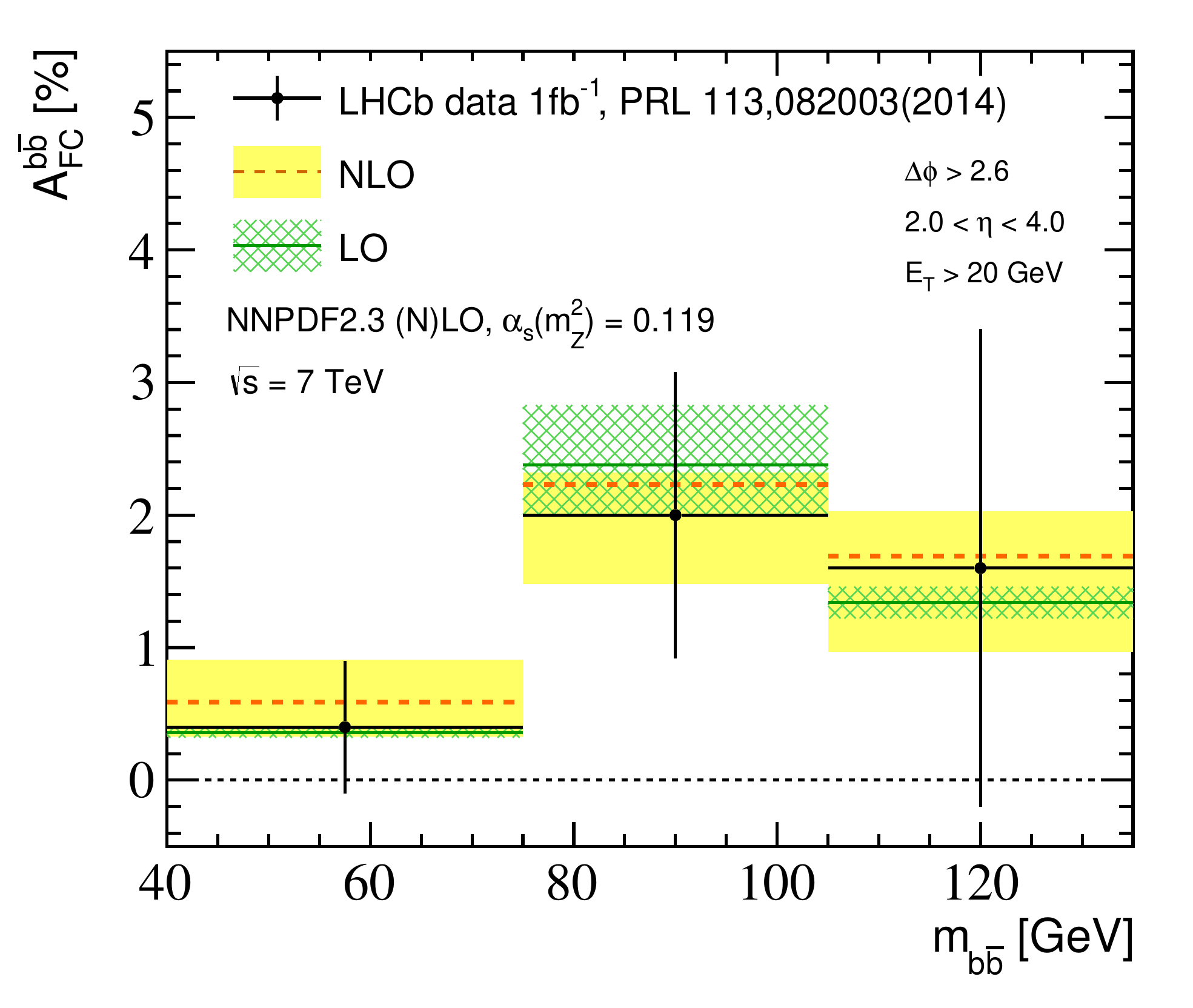}}
\end{center}
\vspace{-8mm}
\caption{NLO and LO predictions of the beauty-quark forward-central asymmetry at $\sqrt{s} =7 \, {\rm TeV}$ within the LHCb acceptance.  The statistical and systematic uncertainties of the measurement have been added in quadrature to obtain the shown experimental error bars. For further details consult the text.}
\label{fig:LHCb7}
\end{figure}

Our NLO and LO predictions for the asymmetry are compared to the existing LHCb data in Fig.~\ref{fig:LHCb7}, and a numerical summary of these predictions is provided in Table~\ref{tab:7TeV}. It is evident that the SM prediction and the data are in full agreement within the given uncertainties, which are currently dominated by the statistical precision of the measurement. Besides this, there are several important features of the theoretical prediction which we would like to mention briefly. 

Similarly to the top-quark asymmetry, the dominant contribution to $A_{\rm FC}^{b \bar b}$ arises from the QCD contribution~$\sigma_a^{s(0)}$ for most values of the invariant mass of the $b$-quark pair, the important exception  being  the central mass bin with $m_{b\bar b} \in [75, 105] \, {\rm GeV}$.  In this region, the double-resonant contribution from $Z$-$Z$ interference becomes dominant, accounting for the bulk of the total asymmetry, as shown in Table~\ref{tab:7TeV}. The QCD correction to this contribution in the resonant bin is observed to slightly decrease the numerator, while the impact on the denominator is marginal. The contribution from mixed QCD-EW and $\gamma$-$Z$ contributions are numerically negligible in this bin (in the case of the QCD-EW corrections this confirms the recent estimate of~\cite{Murphy:2015cha}), which is a consequence of integrating terms that have a single $Z$ propagator over the resonant region. The size of the mixed QCD-EW corrections is further reduced across all bins by a partial cancellation between contributions arising from $u \bar u$ and $d \bar d$ initial states, which is a result of the $u$ and $d$ quarks having opposite electromagnetic/weak charges. We also note that the $qg$-initiated contribution is numerically sizeable, accounting for up to almost 10\% of the entire $\sigma^{s(0)}_a$ term.

\renewcommand*{\arraystretch}{1.5}
\begin{table}[t!]
\centering
  \begin{tabular}{@{} c|c|c|c|c@{}}
			NLO  & $A_{\rm FC}^{b \bar b}$~[\%]& QCD	&QCD-EW & EW  \\ \hline
    			$m_{b \bar b} \in [40, 75] \, {\rm GeV}$ & $0.59 ^{+0.32}_{-0.26}$	& $100.6\%$  & $-4.9\%$ & $4.3\%$ \\ \hline
    			$m_{b \bar b} \in [75, 105] \, {\rm GeV}$ & $2.23 ^{+0.09}_{-0.75} $ & $33.5\%$ & $-1.4\%$ & $67.9\%$ \\ \hline
    			$m_{b \bar b} > 105 \, {\rm GeV}$ & $1.69^{+0.34}_{-0.72}$ & $86.6\%$ & $-7.1\%$ & $20.5\%$ \\ \hline  \hline
			LO & & & & \\ \hline 
			$m_{b \bar b} \in [40, 75] \, {\rm GeV}$ & $0.36 ^{+0.04}_{-0.03}$	& $105.0\%$  & $-5.1\%$ & $0.2\%$ \\ \hline
    			$m_{b \bar b} \in [75, 105] \, {\rm GeV}$ & $2.38 ^{+0.45}_{-0.37} $ & $30.9\%$ & $-1.2\%$ & $70.3\%$ \\ \hline
    			$m_{b \bar b} > 105 \, {\rm GeV}$ & $1.34 ^{+0.12}_{-0.12}$ & $96.8\%$ & $-8.3\%$ & $11.5\%$ 
     \end{tabular}
  \caption{NLO and LO predictions of the $b \bar b$ forward-central asymmetry at $\sqrt{s} =7 \, {\rm TeV}$ within the LHCb acceptance. The relative contributions to $A_{\rm FC}^{b \bar b}$ from QCD, mixed QCD-EW and EW corrections are also provided.}
  \label{tab:7TeV}
\end{table}

The coloured bands around the central values of the~NLO~(yellow) and LO~(green) asymmetry predictions shown in Fig.~\ref{fig:LHCb7} are due to scale variation alone. Away from resonance, the scale uncertainty of the LO asymmetry is artificially small as both numerator and denominator of $A_{\rm FC}^{b \bar b}$ are dominated by the leading QCD contributions $\sigma_a^{s(0)}$ and $\sigma_s^{s(0)}$, respectively. In both cases, the $\mu_R$ dependence enters only via the scale dependence of $\alpha_s$, and two powers of $\alpha_s$ cancel in the ratio. For $m_{b\bar b} \in [75, 105] \, {\rm GeV}$, the LO scale uncertainties are more pronounced as the dominant correction $\sigma_a^{e(0)}$ in that bin has no $\mu_R$ dependence, while the denominator is to first approximation given by $\alpha_s^2 \sigma_s^{s(0)}$. 

Given the unreliable evaluation of theoretical uncertainties in the case of the LO results, and the ambiguity that these results strongly depend on whether NLO or LO PDFs are used for the simultaneous computation of the numerator and denominator, we believe that the NLO predictions are most trustworthy. Let us finally add, that although our NLO and LO predictions are consistent within uncertainties, this feature arises due to the presence of a hard LO gluon PDF at large-$x$ which partially compensates the effect of missing higher-order corrections in the symmetric cross section~\cite{Gauld:2014pxa}. Given that the gluon PDF is essentially decoupled from many observables in a LO global PDF fit, this compensation can be regarded as accidental. Conversely, if the LO predictions are obtained using NLO PDFs, like for instance in~\cite{Murphy:2015cha}, the resulting LO asymmetry predictions tend to be systematically higher by about $50\%$. Further improvement beyond the NLO predictions would require the inclusion of ${\cal O} (\alpha_s^4)$ corrections to both the numerator and denominator in~(\ref{eq:2}), as done in \cite{Czakon:2014xsa} for $t \bar t$ production. Such a computation is beyond the scope of this letter.

\section{\boldmath Predictions for $\sqrt{s} = 13 \, {\rm TeV}$}

In addition to $b \bar b$ asymmetry measurements, LHCb has recently demonstrated the ability to efficiently tag charmed jets~\cite{Aaij:2015yqa}, which in the future may allow for asymmetry measurements involving $c \bar c$ final states, assuming that charge tagging can be established for charm quarks. We therefore also provide predictions for $A_{\rm{FC}}^{c\bar{c}}$ at $\sqrt{s} = 13$~TeV, relevant for data taking in the initial years of LHC Run-II.  

\renewcommand*{\arraystretch}{1.5}
\begin{table}[t!]
\centering
  \begin{tabular}{@{} c|c|c|c|c@{}}
			& $m_{b\bar{b}}$~$[{\rm GeV}]$ & $[40,75]$ & 	$[75,105]$ & $>105$ \\ \hline	
    \multirow{2}{*} {NLO} & $A_{\rm FC}^{b \bar b} \, [\%]$	&  $0.32^{+0.25}_{-0.15}$ & $1.03^{+0.04}_{-0.26}$ 	& $0.80^{+0.17}_{-0.29}$  \\
               & $\sigma_s^{b \bar b} \, [{\rm nb}]$ & $64.2^{+38.1}_{-14.6}$ & $14.0^{+5.4}_{-1.4}$ & $4.18^{+2.15}_{-0.43}$  \\ \hline \hline
    \multirow{2}{*} {LO} & $A_{\rm FC}^{b \bar b} \, [\%]$	&  $0.17^{+0.02}_{-0.02}$ & $1.21^{+0.23}_{-0.19}$ & $0.67^{+0.06}_{-0.05}$  \\
               & $\sigma_s^{b \bar b} \, [{\rm nb}]$ & $118.7 ^{+28.1}_{-20.6}$ & $12.9^{+3.0}_{-2.2}$ & $4.21^{+0.99}_{-0.74}$  \\ 
  \end{tabular}
  \caption{NLO and LO predictions for the $b \bar b$ forward-central asymmetry at $\sqrt{s} =13 \, {\rm TeV}$. Predictions for the symmetric cross section $\sigma_s^{b\bar b}$  within the LHCb acceptance are also given.}
  \label{13TeVBeauty}
\end{table}

Following the $\sqrt{s} = 7 \, {\rm TeV}$ analysis strategy, we provide our fixed-order predictions in bins of the heavy quark-pair invariant mass bins $m_{Q\bar Q}$, and also apply the same $E_T > 20 \, {\rm GeV}$,  $2 < \eta < 4$, and $\Delta \phi > 2.6$ cuts. The total asymmetry and symmetric cross section predictions for the beauty final state are provided in Table~\ref{13TeVBeauty}. The relative contribution to the asymmetry from QCD, mixed QCD-EW and EW corrections is qualitatively unchanged with respect to the $\sqrt{s} = 7 \, {\rm TeV}$ predictions (see~Table~\ref{tab:7TeV}). Importantly, at $\sqrt{s} = 13 \, {\rm TeV}$, the symmetric cross section increases by a factor of  three to five, while the total asymmetry is reduced by approximately a factor of two. This implies that, with $5 \,  {\rm fb}^{-1}$ of integrated luminosity, LHCb should be able to improve the statistical precision of their $b \bar b$ measurement by at least a factor of two. This assumption conservatively assumes no improvement in the tagging efficiency.

\renewcommand*{\arraystretch}{1.5}
\begin{table}[h!]
\centering
 \begin{tabular}{@{} c|c|c|c|c@{}}
			& $m_{c\bar{c}}$~$[{\rm GeV}]$ & $[40,75]$ & 	$[75,105]$ & $>105$ \\ \hline	
    \multirow{2}{*} {NLO} & $A_{\rm FC}^{c \bar c} \, [\%]$	&  $0.40^{+0.34}_{-0.19}$ & $0.78^{+0.09}_{-0.20}$ 	& $0.89^{+0.18}_{-0.31}$  \\
               & $\sigma_s^{c \bar c} \, [{\rm nb}]$ & $57.9^{+40.0}_{-15.4}$ & $13.6^{+5.0}_{-1.2}$ & $4.03^{+2.09}_{-0.38}$  \\ \hline \hline
    \multirow{2}{*} {LO} & $A_{\rm FC}^{c \bar c} \, [\%]$	&  $0.20^{+0.02}_{-0.02}$ & $0.90^{+0.13}_{-0.11}$ & $0.78^{+0.06}_{-0.06}$  \\
               & $\sigma_s^{c \bar c} \, [{\rm nb}]$ & $111.1^{+26.3}_{-19.3}$ & $12.5^{+2.9}_{-2.1}$ & $4.15^{+0.98}_{-0.74}$  \\ 
  \end{tabular}
  \caption{Predictions for charm-quark pair final states  analogue  to Table~\ref{13TeVBeauty}.}
  \label{13TeVCharm}
\end{table}

Our NLO and LO predictions for $c \bar c$ production are summarised in Table~\ref{13TeVCharm}. As pointed out recently in~\cite{Murphy:2015cha}, there is an additional contribution arising at $\mathcal{O}(\alpha_s \alpha)$ due to $s$-channel gluon exchange interfering with $t$-channel $W$-boson. Numerically, this contribution amounts relatively to around $-2\%$ of~$A_{\rm FC}^{c \bar c}$ in the high invariant mass bin, and is insignificant elsewhere. Qualitatively, the charm and beauty asymmetry predictions are similar. However, unlike the beauty predictions the mixed QCD-EW corrections to the asymmetry are positive,  because up-type and down-type quarks have opposite weak charges. This leads to a slightly increased asymmetry in both the low and high invariant mass bins. Another notable difference is that the value of the asymmetry in the resonant bin is reduced for the  charm-quark final state as the magnitude of the vector coupling of the $Z$ boson to up-type quarks is approximately two times smaller than that to down-type quarks.

To better quantify the differences between $A_{\rm{FC}}^{b\bar{b}}$ and~$A_{\rm{FC}}^{c\bar{c}}$, we present predictions for the ratio of these two asymmetries in Table~\ref{13TeVratios}. The differences in these asymmetry ratios due to corrections involving electromagnetic/weak couplings range between $8\%$ to $30\%$, suggesting that an improvement in the experimental systematics will be required to observe such effects --- the systematic uncertainty in the central mass bin of the $\sqrt{s} = 7 \, {\rm TeV}$ asymmetry measurement is about 30\%.

\renewcommand*{\arraystretch}{1.5}
\begin{table}[t!]
\centering
 \begin{tabular}{@{} c|c|c|c|c@{}}
			& $m_{Q\bar{Q}}$~$[{\rm GeV}]$ & $[40,75]$ & 	$[75,105]$ & $>105$ \\ \hline	
    \multirow{2}{*} {NLO} & $A_{\rm FC}^{c \bar c}/A_{\rm FC}^{b \bar b}$	&  $1.25^{+0.04}_{-0.08}$ & $0.76^{+0.09}_{-0.04}$ 	& $1.11^{+0.02}_{-0.03}$  \\
               & $\sigma_s^{c \bar c}/\sigma_s^{b \bar b}$ & $0.90^{+0.05}_{-0.05}$ & $0.98^{+0.03}_{-0.04}$ & $0.97^{+0.05}_{-0.02}$  \\ \hline \hline
    \multirow{2}{*} {LO} & $A_{\rm FC}^{c \bar c}/A_{\rm FC}^{b \bar b}$	&  $1.18^{+0.06}_{-0.01}$ & $0.74^{+0.04}_{-0.02}$ & $1.16^{+0.01}_{-0.02}$  \\
               & $\sigma_s^{c \bar c}/\sigma_s^{b \bar b}$ & $0.94^{+0.00}_{-0.00}$ & $0.97^{+0.00}_{-0.00}$ & $0.99^{+0.00}_{-0.00}$  \\ 
  \end{tabular}
  \caption{Ratios of charm and beauty predictions based on  LHCb acceptance cuts and $\sqrt{s} = 13 \, {\rm TeV}$.}
  \label{13TeVratios}
\end{table}

Another interesting feature of the $b \bar b$ and $c \bar c$ predictions which we wish to highlight is the relative size of symmetric cross section predictions, which is almost entirely due to QCD. Although the scale uncertainties are still significant at $\mathcal{O}(\alpha_s^3)$, the ratio of charm and beauty cross sections is, for large values of the invariant mass of the heavy-quark pair, expected to be robust with respect to higher-order QCD corrections. We display our NLO and LO predictions for the ratio $\sigma_s^{c \bar c}/\sigma_s^{b \bar b}$ of symmetric cross sections in Table~\ref{13TeVratios}.\footnote{At LO the scale uncertainties almost perfectly cancel between the numerator and denominator of $\sigma_s^{c \bar c}/\sigma_s^{b \bar b}$, so that the uncertainties shown in Table~\ref{13TeVratios} for this case are not a reliable estimate of higher-order corrections.} Given that this observable is theoretically under control, and directly sensitive to charm-tagging and beauty-tagging efficiencies, it should prove useful for validating these efficiencies experimentally for large values of $m_{Q\bar Q}$.

\section{Conclusions}

In this letter we have  presented the results of an improved fixed-order computation for the forward-central asymmetry~$A_{\rm{FC}}^{b\bar{b}}$ in beauty-quark pair production, including a realistic evaluation of scale uncertainties. Our NLO predictions agree within uncertainties with the recent LHCb measurement of this quantity that has been performed with~$1 \, {\rm fb}^{-1}$ of $\sqrt{s} = 7 \, {\rm TeV}$ data. Anticipating improved LHCb measurements of the $b \bar b$ asymmetry, and a first measurement of its counterpart involving charm-quark pairs, we have additionally provided predictions for $A_{\rm{FC}}^{b\bar{b}}$ and $A_{\rm{FC}}^{c\bar{c}}$ at $\sqrt{s} = 13 \, {\rm TeV}$.

Another interesting observable, which we have  identified, is the ratio of the differential $c \bar c$ and $b \bar b$ production cross sections at high invariant mass. Unlike many heavy quark-pair observables, this ratio is expected to be robust with respect to higher-order QCD corrections. This feature can be used to validate charm-jet tagging efficiencies and mis-tag rates in future LHCb analyses. A very good understanding of these experimental issues is a prerequisite for any attempt to search for processes such as $pp \to  h \, (\to  c\bar{c}) \, W,Z$  in the forward region.

\section{Acknowledgements}

We are grateful to Kostas~Petridis and Mike~Williams for several discussions regarding the LHCb measurement of $A_{\rm FC}^{b \bar b}$ at $\sqrt{s} = 7 \, {\rm TeV}$ and thank  Guido~Bell and Christopher~Murphy for fruitful conversations. UH acknowledges the hospitality and support of the CERN theory division.

\end{document}